\documentclass[]{article}
\usepackage{booktabs} 

\begin{document}
\title{Location data quality in context: directions and challenges} 
\author{Maria Luisa Damiani\\  University of Milan, Italy\\ E-mail: maria.damiani@unimi.it}
\date{}

%
%


%
%


\maketitle



\section{Real-time location}
In the last decade, following the emergence of the mobile application domain, the  significance of location information has changed radically. Nowadays, location data  not only is a key component of geospatial databases, 
but also  a critical resource for a broad spectrum of  applications, including Location-based Services and IOT solutions. 
This change of perspective has been made possible by the availability of localization technologies providing 
accurate and reliable location information on moving entities in real time. We refer to real-time location as \emph{positioning} data. 
Positioning data has unique characteristics, such as the strong dependency from both the localization technology and the environmental context,   
which make the specification of standard quality metrics and quality assessment procedures a complex task. 
In this paper, we elaborate on such an aspect, focusing in particular on indoor positioning data. \\
\\
\noindent
\textbf{Positioning data.} In general, positioning data is obtained from the processing of signals -  often radio signals - emitted by a transmitting source and typically propagated along a direct path, i.e. line-of-sight, to a (mobile) receiver. 
The Global Positioning System (GPS) is the first system that  has shown the benefits of accurate location information in real time, revolutioning the field of outdoor navigation. The key of the technology is to
measure the  time elapsed for a radio signal to travel from satellites to a mobile device (called Time-of-Arrival, TOA). The computational mechanism of triangulation is then used to compute the location based on the known position of satellites and TOA-based distance measures \cite{Gentile2013}.   
The system, however, lacks coverage
indoor and in urban areas where physical obstacles block the signal. 
Another popular class of positioning systems  leverages the  Wi-Fi access points (AP) spread across urban/indoor environments through the use of RSS- fingerprinting, a method alternative to TOA.  Such a fingerprint
is defined by  a vector of signal identity information (such as the AP identifier) and a corresponding vector of RSS values  \cite{Ion2014}. 
A successful deployment of the technique in the urban/indoor context has been  pioneered by  Skyhook Wireless and then endorsed by the major players in the mobile device industry such as Apple and
Google. 
In these system, the association between Wi-Fi signatures and locations is stored in  a database. Hence, 
upon a location request, the mobile
device compares the RSSs measured from the APs nearby to those in the
database using a pattern matching technique to detect the best matching \cite{Gentile2013}. 
The technique delivers 
accuracy of tens of meters. However these systems are dependent on the network infrastructure.\\

\noindent 
\textbf{Positioning data quality.}
In general the quality of positioning data 
is tightly dependent on the capabilities or \emph{performance} of the localization system used. In fact, the location information is acquired in real-time, upon request, therefore location is consumed 
as it is computed, without filters in between.  To further clarify this aspect, consider, for example, location accuracy, which represents a major characteristic not only of positioning data  but also of mapping data \cite{Batini2016}. If the location is acquired in real time, its accuracy is necessarily upper bounded by the accuracy of the localization system used to detect it; conversely, in traditional mapping applications, 
the accuracy results from a complex and time consuming map construction process conducted off-line. In this sense the localization technology can be seen as one of the  dimensions of positioning data. 

Data quality is also extremely sensible to  %
the environmental context of the location. Physical objects such as  buildings and walls, influence signal propagation causing, in particular, 
\emph{multi-path} effects (i.e. the reception of multiple copies of the transmitted signal) and \emph{non-line-of-sight}, NLOS, conditions (i.e. obstruction of the direct propagation line between the transmitter and the receiver), which cause signal degradation \cite{Gentile2013}.  For example, in systems based on RSS-fingerprinting,  multi-path causes the fluctuation of the signal strength at a location; while  NLOS causes the random attenuation of the signal strength at different locations in the same area \cite{Gentile2013}. As a result, the  quality delivered at run-time varies in time, depending on both the characteristics of the signal, the localization technique and the physical space.
The impact of  the environmental context is  especially critical  
indoor, where the physical space is complex,  made of rooms, hallways, floors, and populated by objects obstructing the signal and moving people.
The market potential for indoor positioning is however impressive, estimated of several billion dollars by 2022. Application domains include, e.g., retailing and e-marketing, emergency response, assisted living, personal navigation, smart buildings. In a recent report Deloitte  includes indoor positioning among the driving technologies for the next 6 years \cite{Deloitte2017}. In such a context, key question is how to define and evaluate the performance of indoor localization systems and thus,  indirectly, the quality of positioning data.


\section{Indoor localization and evaluation metrics}
\textbf{Localization technologies.} We briefly review major trends. The Wi-Fi (802.11) fingerprinting is  currently the most popular indoor localization technique. Nevertheless, the Wi-Fi network infrastructure is not designed for the localization task,
moreover the fingerprints can deteriorate in time, because the network infrastructure evolves, 
while the level of accuracy may be not sufficient 
\cite{Ion2014}. Complementary  technologies, most at early stage of development, are briefly described next. These technologies are  
 based on radio-frequency, sensors and LED lighting. 
\begin{itemize}
\item 
\emph{Radio frequency based systems.}
	Bluetooth Low Energy (BLE) is a new standard developed by
	Bluetooth Special Interest Group that has attracted the interest of major players e.g. Apple (IBeacon) and Google (Eddystone platform). 
	The signal operates in the same bandwith used by Wi-Fi, while the signal reachable distance is shorter, because of the low power usage policy \cite{Ion2014}.
	A different class of solutions are those based on ultra-wide-band technology (UWB).  UWB-solutions utilizes a large bandwidth which reduces the signal deterioration and thus has the
	potential for very accurate distance estimation.  

\item 
\emph{LED light based systems.} The technology employs the light impulses  emitted by a LED source to communicate the source identifier to a receiver. 
This information is then used along with signal characteristics (e.g. TOA, RSS) to estimate the location \cite{CSUR2015}.
\item \emph{Magnetic field mapping.} This is a fingerprint based technique, which exploits the distortion of the magnetic field due to building materials to assign locations a magnetic signature \cite{Sens2011}. 
\item \emph{Inertial sensors based systems.} Inertial sensors such as accelerometers
and gyroscopes, available on mobile phones, are used to quantify changes in speed or direction. Such information can be used to detect the location based on the starting location and orientation.  This technique is often used in conjunction with other localization methods.
\end{itemize}
\noindent
\textbf{Performance metrics.} The heterogeneity of indoor localization systems and the diversity of application requirements have hindered the development of a common set of concepts and procedures for evaluating the performance of the different localization systems \cite{Survey2009}.  In fact, evaluation methods are in most cases proposed for specific technologies, e.g. \cite{Ion2014,Haute2016}. 
In general, the capabilities of a system are evaluated with respect to a set of \emph{performance metrics.} 
For example, the early framework presented in \cite{Survey2007} introduces the following dimensions for the evaluation of indoor localization system performance:\emph{ accuracy, precision,
complexity, scalability, robustness, and cost}.

Notably, the recent standard ISO/IEC 18305 'Test and evaluation of localization and tracking systems', promotes the use of common practices for the evaluation of localization systems in the large \cite{ISO2016}. The standard introduces detailed metrics primarily for the evaluation of localization accuracy. Location can be specified at two different granularities, either in terms of coordinates or  zones/floors.   Additional metrics include: \emph{coverage}, a measure of the space in which the system meets the minimum performance requirements specified by the application; \emph{latency}, the time elapsed to deliver the location estimate to the ultimate user of the location information: \emph{set-up time}, the time elapsed to make the system operational. Besides, 'optional' performance metrics are specified, such as time availability and resilience for use in mission critical applications. Data security and privacy are included as system requirements.

\noindent
\section{Challenges and directions}
We are witnessing a rapid evolution of the indoor localization techniques. 
For the effective utilization of positioning data a number of challenging issues are to be solved, including:  

\noindent 
\textbf{Increased accuracy}. Centimeter-level accuracy, also referred to as \emph{micro-location}, is the new frontier \cite{Zafari2016}. This calls for advanced methods capable of mitigating the effect of the environmental context on signal propagation. For example,  UWB-based positioning has the potential to match the accuracy requirements, nevertheless such a  performance can be typically achieved in  environments in which there is no physical obstruction because the signal is sensitive to NLOS problems \cite{Gentile2013}, thus methods are needed to mitigate this effect, e.g. \cite{UWB2014}.  
	Another major direction is developing algorithms to fuse sensor data from multiple sources to produce improved localization  \cite{Gentile2013,Zafari2016}.

\noindent
\textbf{Security and privacy}. As the location information becomes a critical resource, preserving the integrity, availability and confidentiality of the geolocation system against cyber attacks becomes essential. Attacks can cause the delivery of incorrect positioning data, system operation disruption, unauthorized location and personal identity disclosure to third parties. Location authentication methods are needed to accomplish critical location-dependent tasks such as the enforcement of context-based access control policies \cite{Damiani2007}.  
	
\noindent
\textbf{Evaluation practices}. The homogenization of performance assessment practices is fundamental to enable the comparison of research  results, typically  evaluated in different environments using different evaluation criteria \cite{Haute2016}.  In that respect, the standardization efforts recently  put in place for the evaluation of localization techniques is a relevant step ahead.  Maintaining the performance metrics and assessment procedures in line with the technological evolution and the application requirements is a major issue.\\	
\\
\emph{	
The work reported in this paper was carried out while visiting the research group of Professor Elisa Bertino at Purdue University CS Department and has been partially supported by the NSF award IIS-1636891.}


\end{document}